\documentclass[letter]{jpsj2} 

\title
{Change of Electronic Structure Induced by Magnetic Transitions in CeBi}

\author
{
Shin-ichi {\sc Kimura}$^{1,}$\thanks{E-mail: kimura@ims.ac.jp},
Mitsuru {\sc Okuno}$^{2,}$\thanks{Present address: LSI Laboratory, Mitsubishi Electric Corporation, Itami, Hyogo},
Hideaki {\sc Kitazawa}$^{3}$,
Fumihiko {\sc Ishiyama}$^{4}$
 and Osamu {\sc Sakai}$^{5}$
}

\inst
{
$^1$UVSOR Facility, Institute for Molecular Science, Okazaki 444-8585\\
$^2$Graduate School of Science and Technology, Kobe University, Nada-ku, Kobe 657-8501\\
$^3$Nanomaterials Laboratory, National Institute for Materials Science, Tsukuba, 305-0047\\
$^4$NTT Energy and Environment Systems Laboratories, Tokyo 180-8585\\
$^5$Department of Physics, Tokyo Metropolitan University, Hachioji, Tokyo 192-0397
}

\recdate
{
February 18, 2004
}

\abst
{
The temperature dependence of the electronic structure of CeBi arising from two types of antiferromagnetic transitions based on optical conductivity ($\sigma(\omega)$) was observed.
The $\sigma(\omega)$ spectrum continuously and discontinuously changes at 25 and 11~K, respectively.
Between these temperatures, two peaks in the spectrum rapidly shift to the opposite energy sides as the temperature changes.
Through a comparison with the band calculation as well as with the theoretical $\sigma(\omega)$ spectrum, this peak shift was explained by the energy shift of the Bi $6p$ band due to the mixing effect between the Ce $4f \Gamma_8$ and Bi $6p$ states.
The single-layer antiferromagnetic ($+-$) transition from the paramagnetic state was concluded to be of the second order.
The marked changes in the $\sigma(\omega)$ spectrum at 11~K, however, indicated the change in the electronic structure was due to a first-order-like magnetic transition from a single-layer to a double-layer ($++--$) antiferromagnetic phase.
}

\kword
{
CeBi, optical conductivity, $pf$ mixing
}

\begin{document}
\sloppy
\maketitle

Materials composed of rare earth compounds have many attractive physical properties due to the alternately localized and itinerant nature of their $4f$ electrons.
Among the rare-earth compounds, cerium monopnictides (Ce$X$; $X$ = P, As, Sb and Bi) are known as low-carrier-density systems.
Since they have various physical properties different from ordinary dense Kondo systems, many experimental and theoretical studies have been performed in an attempt to learn more about their electronic behavior.
The most significant physical property of Ce$X$ compounds is the complicated magnetic phase diagrams and associated magnetic structures that appear at low temperatures, high magnetic fields, and high pressures.~\cite{RM-CeSb,Kohgi1}
The main origin of these complex structures has been qualitatively attributed to be the hybridization between the Ce $4f$ and the $X$ $p$ states, the so-called $pf$ mixing.~\cite{TakahashiKasuya}
In CeSb, Kimura and coworkers, and Ishiyama and Sakai pointed out that not only the $pf$ mixing but also the hybridization between Ce $5d$ and $X p$ states occurs primarily in the ordered states on the basis of a comparison of theoretical calculations with magneto-optical spectra.~\cite{Kimura-CeSb1, Kimura-CeSb2, IshiyamaSakai}

The typical structure of the magnetically ordered states of Ce$X$ is that of double ferromagnetic layers.
For instance, the antiferromagnetic phase of CeSb below 9.5~K is $++--$, where $+ (-)$ indicates an up (down) spin layer along the magnetic field.~\cite{RM-CeSb}
CeBi has the same magnetic structure, referred to as the AF-1A phase, below 11~K at zero magnetic field.~\cite{RM-CeBi}
However, CeBi has a unique magnetic structure, that of a single-layer antiferromagnetic phase ($+-$) referred to as AF-1, which is not present in CeSb at ambient pressure and only exists at temperatures between 11 and 25~K in CeBi, as revealed by elastic neutron scattering data.~\cite{Osakabe}
The lattice periodicity due to the spin and/or orbital structure changes from that in the paramagnetic state upon the phase transition of AF-1.
The periodicity can be detected by X-ray diffraction and elastic neutron scattering.~\cite{Osakabe}
If the band structure couples to the spin and/or orbital structure, the band structure will be folded by the periodicity.
Then, changes in the electronic structure can be observed by optical conductivity ($\sigma(\omega)$) and angle-resolved photoemission measurements.~\cite{Ito-CeSb}
Pittini and coworkers have indicated that the folding of the band structure of CeBi by the spin structure can be observed in the $\sigma(\omega)$ spectra.~\cite{Pittini}
Here, the spin moments originating from the Ce $4f^1 \Gamma_8$ and $\sigma(\omega)$ spectra reflect the Bi $6p$ and Ce $5d$ band structure.
The coupling between the Ce $4f^1 \Gamma_8$ spin moment and the band structure is due to $pf$ mixing.
The $pf$ mixing intensity is constant in each magnetically ordered state in CeSb because the $\sigma(\omega$) spectra as well as the corresponding electronic structure do not change in any of the phases.~\cite{Kimura-CeSb1,Kimura-CeSb2}
However, in CeBi, we found that the $\sigma(\omega$) spectrum in the AF-1 phase gradually changes with temperature.
The origin of this change is the topic of this Letter.

CeBi was produced using the Bridgman method with a tungsten heater.~\cite{Suzuki-Sample}
The sample with a size of 4~mm in diameter and 1~mm in thickness was cleaved along the (001) plane in a helium atmosphere and was placed in a closed-cycle-helium cryostat in situ to avoid oxidation.
The dependence of the reflectivity spectrum on temperature down to 6~K was measured mainly in the photon energy range of 10~meV - 1.5~eV using a conventional Fourier-transform interferometer (JASCO FTIR610) combined with custom-made high vacuum optics using a cryostat for sample cooling.
The temperature dependence measurement was performed at temperatures from 7 to 30~K in 1~K steps.
The base pressure of the sample chamber was less than $5 \times 10^{-7}$~Pa to avoid ice contamination at low temperatures.
The reflectivity spectrum up to $\hbar \omega$ = 200~eV was acquired at UVSOR for the Kramers-Kronig analysis.~\cite{Fukui-BL7B}
The high-energy spectrum was connected to the reflectivity spectra at low temperatures in the energy range of 0.01 - 2~eV to obtain the $\sigma(\omega$) spectra.
The reflectivity spectrum above 2~eV does not significantly change with temperature.
In the energy range below 0.01~eV and above 200~eV, the spectra were extrapolated using the Hagen-Rubens function and $R(\omega) \propto \omega^{-4}$, respectively.~\cite{Wooten}

\begin{figure}[t]
\begin{center}
\includegraphics[width=8cm]{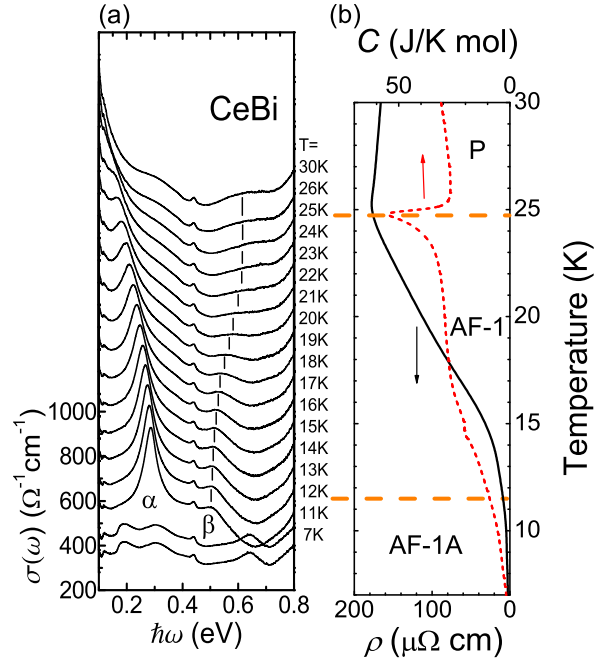}
\end{center}
\caption{(a) Optical conductivity spectra ($\sigma(\omega)$) of CeBi between 7 and 30~K.
The temperature dependence of the $\alpha$ and $\beta$ peaks are discussed in the text.
(b) The electrical resistivity ($\rho$) and the specific heat ($C$) in the same temperature region.
The horizontal dashed lines indicate the magnetic transition temperatures.
}
\label{OC-fig}
\end{figure}
The obtained $\sigma(\omega)$ spectra are plotted in Fig.~\ref{OC-fig}~(a) and the corresponding electrical resistivity ($\rho$)~\cite{TK} and specific heat ($C$)~\cite{TS} data are indicated in Fig.~\ref{OC-fig}~(b).
The $\rho$ and $C$ data have a small anomaly (does not appear in this figure) at 11~K and a large peak at 25~K.
The former and latter phase transitions are known to originate from the magnetic phase transitions from the AF-1A phase $(++--)$ to the AF-1 phases and from the AF-1 to paramagnetic (P) phases, respectively.
Considering only the magnetic structure changes in the transition from AF-1 to AF-1A at 11~K, the $\rho$ and $C$ data may not have a large anomaly.
On the other hand, the $\sigma(\omega)$ spectrum discontinuously and continuously changes at 11 and 25~K, respectively.
This behavior is opposite to that of the $\rho$ and $C$ results.
Therefore, the magnetic transition from AF-1 to AF-1A is accompanied by a change in the electronic structure that contributes to the $\sigma(\omega)$ spectrum.
On the other hand, the magnetic transition from P to AF-1 occurs without a change in the electronic structure.
On the basis of these observations, the effect of the magnetic transition from P to AF-1 on the electronic structure is different from that from AF-1 to AF-1A.

In the AF-1 phase, two peaks, referred to as $\alpha$ and $\beta$ at 0.3 and 0.5~eV, at 12~K in Fig.~\ref{OC-fig} (b) shift to lower and higher energy values, respectively, with increasing temperature.
Their intensities decreased as temperature increased until there was no significant peak structure in the P phase.
The total energies of the peak shifts seem to be approximately equal even though these peaks shift in opposite directions energetically.
This behavior indicates these peaks must have a common origin.

The change in the electronic structure and in the $\sigma(\omega)$ spectrum from the AF-1 to AF-1A transition originates from the change in the magnetic structure via the $pf$ mixing, judging from the past results of CeSb.~\cite{Kimura-CeSb1, Kimura-CeSb2} 
Pittini and coworkers pointed out that the electronic band structure is folded by the spin structure in the AF-1 phase.~\cite{Pittini}
As previously mentioned, the change in the $\sigma(\omega)$ spectrum can be explained by the band folding.
However, the change in the $\sigma(\omega)$ spectrum between 11 and 25~K has never been previously observed.
In addition, the $\sigma(\omega)$ spectrum of the AF-1 phase smoothly connects to that of the P phase.
These results cannot be explained only on the basis of the band folding.
Here, one Bi $6p$ band of the four degenerated bands at the $\Gamma$ point in the P phase should be pushed up by the occupied $4f$ state to increase the $pf$ mixing, as observed in CeSb in the ferromagnetic phase.~\cite{Kimura-CeSb1}

\begin{figure}[t]
\begin{center}
\includegraphics[width=7cm]{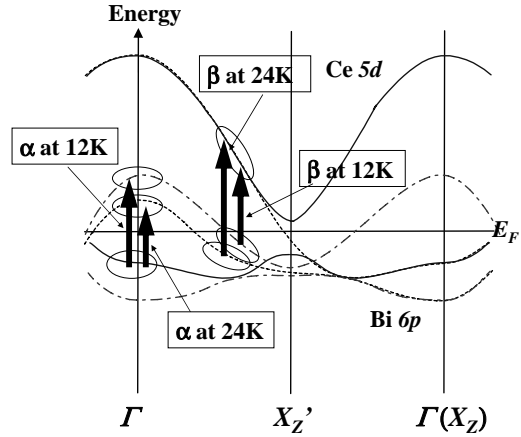}
\end{center}
\caption{
Predicted band structure and optical transitions of CeBi at 12 and 24~K in AF-1 phase.
The dotted line is the band structure in the P phase and that of nonmagnetic LaBi.~\cite{Hasegawa85}
The solid and dot-dashed lines mainly originate from Ce $5d$ states and Bi $6p$ states, respectively.
The band structure is folded along the $\Gamma - X_Z$ axis by the $+ -$ spin structure due to the presence of $pf$ mixing.
However, the energy band does not change from that in the P phase in the case of the weak $pf$ mixing near 25~K.
The $pf$ mixing also pushes up one of the Bi $6p$ bands that then mixes with the Ce $4f \Gamma_8$.
This effect also appears at the $\Gamma$-point.
}
\label{band-fig}
\end{figure}
The predicted electronic structure of CeBi in the AF-1 phase is shown in Fig.~\ref{band-fig}.
The band structure in the P phase refers to that of LaBi.~\cite{Hasegawa85}
One of the Bi $6p$ bands having the same symmetry as the Ce $4f \Gamma_8$ band is pushed up by the $pf$ mixing.
This situation also exists in CeSb.~\cite{Kimura-CeSb2}
Since the AF-1 phase spin structure is $+ -$, the periodicity should be twice that of the fundamental lattice structure.
The energy band should then be folded into twice that of the P phase.
The $pf$ mixing, then, operates not only to fold the energy band but also to push up the Bi $6p$ state.
This band structure of the AF-1 phase is represented by both the solid and dot-dashed lines in Fig.~\ref{band-fig}.

In this model, the $\sigma(\omega)$ peaks $\alpha$ and $\beta$ are considered to originate from the transitions of Ce $5d$ $\rightarrow$ Bi $6p$ near the $\Gamma$ point and of Bi $6p$ $\rightarrow$ Ce $5d$ near the $\Gamma - X_Z$ axis as shown in Fig.~\ref{band-fig}.
The former transition may occur around the high symmetry point because the transition probability of Ce $5d \rightarrow$ Bi $6p$ is generated through the mixing due to the band folding.
The peak originates from the $M_1$ critical point with the van Hove singularity inferred from the peak shape.~\cite{DresselGruner}
At 12~K, the $pf$ mixing is present, resulting in the Bi $5p$ band being pushed up.
With increasing temperature, the $\sigma(\omega)$ peaks ($\alpha$ and $\beta$) shift to the low- and high- energy sides, followed by a decrease in their intensities with increasing temperature.
Finally, the $\sigma(\omega)$ spectrum smoothly transforms to the P phase.
This indicates that the $pf$ mixing intensity ($I_{pf}$) decreases with increasing temperature.
As $I_{pf}$ decreases, the Bi $6p$ band moves towards a low energy, and the $\alpha$ and $\beta$ peaks shift to the low- and high-energy sides, respectively.
In addition, the folded Ce $5d$ band near the $\Gamma$ point gradually disappears.
This is followed by a gradual decrease in the intensity of the $\alpha$ peak.
In the case of $\beta$, the peak shifts to the high-energy side while broadening, but remains visible even in the P phase because the Bi $6p$ peak only shifts to the lower-energy side with increasing temperature at around the $\Gamma - X_Z$ axis.

The change in $I_{pf}$ suggests that the phase transition from P to AF-1 is second order.
This is the basis of the $\sigma(\omega)$ spectrum in the AF-1 phase continuously transforming into that of the P phase.
Both the temperature-dependent lattice constant~\cite{Hulliger} and the thermal expansion data~\cite{Sera} suggest the same conclusion.
In addition, Takahashi and Kasuya predicted that the phase transition would be of the second order due to the short-range order.~\cite{TakahashiKasuya}

\begin{figure}[t]
\begin{center}
\includegraphics[width=6cm]{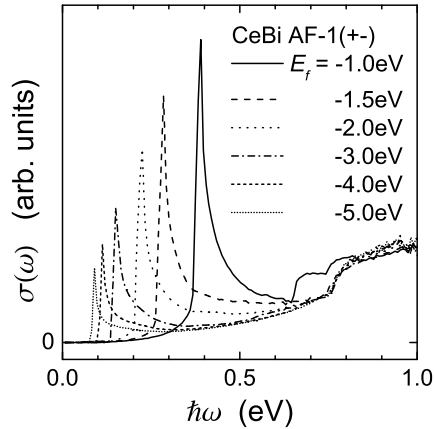}
\end{center}
\caption{
Calculated $\sigma(\omega)$ spectra of CeBi in AF-1 phase as a function of effective Ce $4f$ level ($E_f$).
The $pf$ mixing intensity is almost proportional to $-1/E_f$.
}
\label{calculatedOC}
\end{figure}
To confirm that the temperature dependence of the $\sigma(\omega)$ spectrum can be attributed to a second-order transition, the $\sigma(\omega)$ spectra in the AF-1 phase with different $I_{pf}$ were calculated as shown in Fig.~\ref{calculatedOC}.
In this calculation, we employed the tight binding band model with the $pf$ mixing, which was used to obtain $\sigma(\omega)$ of CeSb.~\cite{Kimura-CeSb2, IshiyamaSakai}
In this model, we retained the bands in the Hamiltonian in which the $pf$ mixing in the magnetically ordered states has a strong effect, and we treated the other bands, which have Fermi surfaces, as reservoir bands.
In addition, we introduced the Schrieffer-Wolff transformation~\cite{Schrieffer} to avoid treating the $f$ states as $f$ bands. 
With this transformation, the $pf$ mixing term is replaced by the effective $p - p$ term, and $I_{pf}$ is controlled by changing the effective $f$ level ($E_f$), measured from the Fermi energy of the paramagnetic phase.
The band parameters were fitted so as to reproduce the band structure of LaBi in the paramagnetic phase.~\cite{Hasegawa85}

The significant peak corresponds to the $\alpha$ peak in Fig.~\ref{OC-fig}.
The $\beta$ peak does not appear in Fig.~\ref{calculatedOC}.
The reason for this is that the band calculation of LaBi is not suitable for the complete explanation of the $\sigma(\omega)$ peaks.
A full band calculation, for instance, like the one by Kaneta and coworkers~\cite{Kaneta}, should be used for more detailed discussions.

$I_{pf}$ is almost proportional to $-1/E_f$.~\cite{IshiyamaSakai}
Figure~\ref{calculatedOC} indicates that both the peak energy and the intensity decrease with increasing $-E_f$ as well as with decreasing $I_{pf}$.
This implies the temperature dependence of the $\alpha$ peak can be represented by the change in $I_{pf}$.

\begin{figure}[t]
\begin{center}
\includegraphics[width=7cm]{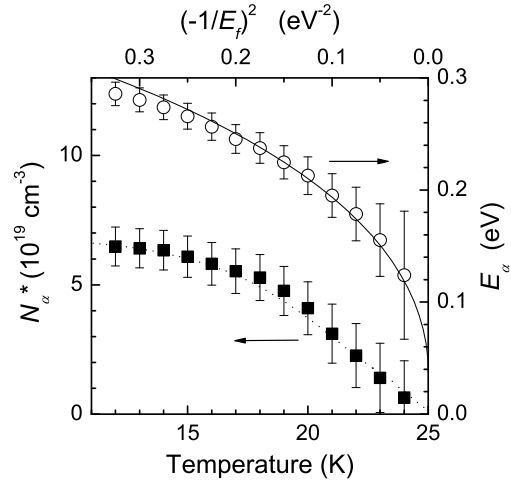}
\end{center}
\caption{
Effective electron number ($N_\alpha^*$, solid square) and energy ($E_\alpha$, open circle) of $\alpha$ peak in Fig.~\ref{OC-fig} as functions of temperature.
The dotted and solid lines are $N_\alpha^*$ and $E_\alpha$, respectively, derived from the calculated $\sigma(\omega)$ in Fig.~\ref{calculatedOC} as functions of the Ce $4f$ level ($E_f$).
See the text for details.
}
\label{intensity-fig}
\end{figure}

To derive the relationship between $I_{pf}$ and temperature, the energy ($E_\alpha$) and the effective electron number ($N_\alpha^*$) of the $\alpha$ peak were compared with the theoretical values.
$N_\alpha^*$ is obtained by
\[
N_\alpha^* = \frac{2 m_0}{\pi e^2} \int_0^\infty \sigma_\alpha(\omega) d\omega.
\]
Here, $m_0$ and $e$ are the bare mass and the charge of an electron, respectively, and $\sigma_\alpha(\omega)$ is the $\alpha$ peak after subtracting the background, mainly derived from the carriers' absorption.
The experimentally obtained $E_\alpha$ and $N_\alpha^*$ are plotted with open circles and solid squares, respectively, as functions of temperature, as shown in Fig.~\ref{intensity-fig}.
The error bars were ascribed to the error in the background subtraction.

In the case of the second-order transition, $I_{pf}$ should be proportional to $(T_{\rm N}-T)^{1/2}$, where $T_{\rm N}$ is the N\'eel temperature of 25~K.
The theoretical $E_\alpha$ and $N_\alpha^*$ as functions of $E_f$ can be obtained from Fig.~\ref{calculatedOC}.
The theoretical $E_\alpha$ and $N_\alpha^*$ were derived from the fitting of the peak energy and integration of the peak subtracted from the background originating from the tail of the higher absorption part of the spectrum, respectively.
Since $T_{\rm N}-T \propto I_{pf}^2 \propto (-1/E_f)^2$, $E_\alpha$ and $N_\alpha^*$ can be written as
\[
E_\alpha = \Delta E_\alpha(T_{\rm N}-T) + E_{\alpha 0} = \Delta E_\alpha((-1/E_f)^2)+E_{\alpha 0}
\]
\[
N_\alpha^* = N_\alpha^*(T_{\rm N}-T)=N_\alpha^*((-1/E_f)^2).
\]
Here, $\Delta E_\alpha(T)$ is the temperature-dependent part of the peak energy and $E_{\alpha 0}$ is the optical transition energy at $I_{pf}=0$.
The theoretical $E_\alpha$ and $N_\alpha^*$ as functions of $(-1/E_f)^2$ are plotted by solid and dashed lines, respectively, in Fig.~\ref{intensity-fig}.
Both $E_\alpha$ and $N_\alpha^*$ were fitted by the above functions accurately.
$E_{\alpha 0}$ was found to be 0.05~eV on the basis of the fitting.
The relationship between $T$ and $E_f$ was found to be $T=25-42.4/E_f^2$.
This demonstrates that the behavior of the $\alpha$ peak can be explained by a second-order antiferromagnetic transition.

At a low temperature, the phase changes from AF-1 to AF-1A at 11~K.
This temperature corresponds to $(-1/E_f)^2 = 0.33$~eV$^{-2}$ ($E_f \sim -1.74$~eV).
Therefore, the double-ferromagnetic-layer structure is stable at $E_f \geq$ -1.74~eV.
The origin of the double-layer structure in AF-1A has been discussed on the basis of the $pf+pd$ mixing model proposed by Ishiyama and Sakai.~\cite{IshiyamaSakai}
In this model, the double-ferromagnetic-layer structure stabilizes at $E_f \geq$ -1.7~eV.
This shows that the phase transition can indeed be quantitatively explained by the $pf+pd$ mixing model.

To summarize, we noted two $\sigma(\omega)$ peaks with strong temperature dependencies that originate from the transitions between the Bi $6p$ and Ce $5d$ states of CeBi in the AF-1 phase.
The origin of the temperature dependence is the onset of $pf$ mixing with decreasing temperature.
The transition from the P to AF-1 phases was concluded to be a second-order one.
This is the first observation of a second-order magnetic transition clearly appearing in an optical spectrum.
On the other hand, in the transition from AF-1 to AF-1A, the $\sigma(\omega)$ spectrum as well as the band structure markedly change as a result of the different magnetic structures of the phases.
The transition temperature as well as the $pf$ mixing intensity can be qualitatively explained by the $pf+pd$ mixing model as proposed by Ishiyama and Sakai.

This work was a joint study program of the Institute for Molecular Science and partially supported by a Grant-in-Aid for Scientific Research from MEXT of Japan.
Numerical calculations were partially performed at the Institute of Scientific and Industrial Research, Osaka University.
One of the authors (S.K.) would like to thank Prof. Kwon for fruitful discussion.


\end{document}